\documentclass[conference]{IEEEtran}
\usepackage[letterpaper, left=0.65in, right=0.65in, bottom=1in, top=0.7in]{geometry}
\usepackage{cite,microtype}
\usepackage{amsmath,amssymb,amsfonts}
\usepackage[subtle]{savetrees}
\usepackage{algorithmic}
\usepackage{subcaption}
\usepackage{mathtools}
\usepackage{tikz}
\usepackage{graphicx}
\usepackage{textcomp}
\usepackage{layout}
\DeclareMathOperator*{\argmax}{arg\,max}

\usepackage{xcolor}
\def\BibTeX{{\rm B\kern-.05em{\sc i\kern-.025em b}\kern-.08em
    T\kern-.1667em\lower.7ex\hbox{E}\kern-.125emX}}

\title{Team Deep Mixture of Experts for Distributed Power Control
\thanks{The work of M.Zecchin is founded by Marie Skłodowska-Curie actions
	(MSCA-ITN-ETN 813999 WINDMILL)}}

\author{\IEEEauthorblockN{Matteo Zecchin,
		David Gesbert, Marios Kountouris}
	\IEEEauthorblockA{Communication Systems Department,
		EURECOM\\
		Biot, France\\
		Email:\{ zecchin,
		gesbert,
		kountour\}@eurecom.fr}}
\begin{document}
	\maketitle
	\begin{abstract}
		In the context of wireless networking, it was recently shown that multiple DNNs can be jointly trained to offer a desired collaborative behaviour capable of coping with a broad range of sensing uncertainties. In particular, it was established that DNNs can be used to derive policies that are robust with respect to the information noise statistic affecting the local information (e.g. CSI in a wireless network) used by each agent (e.g. transmitter) to make its decision. While promising, a major challenge in the implementation of such method is that information noise statistics may differ from agent to agent and, more importantly, that such statistics may not be available at the time of training or may evolve over time, making burdensome retraining necessary. This situation makes it desirable to devise a “universal" machine learning model, which can be trained once for all so as to allow for decentralized cooperation in any future feedback noise environment. With this goal in mind, we propose an architecture inspired from the 	well-known Mixture of Experts (MoE) model, which was previously used for non-linear regression and classification tasks in various contexts, such as computer vision and speech recognition. We consider the decentralized power control problem as an example to showcase the validity of the proposed model and to compare it against other power control algorithms. We show the ability of the so called Team-DMoE model to efficiently track time-varying statistical scenarios.
	\end{abstract}
	
	\begin{IEEEkeywords}
		team decision theory,  machine learning, distributed power control, interference channel, wireless network
	\end{IEEEkeywords}
	\section{Introduction}
	The potential gains coming from cooperative behaviour between wireless transmitters has been been firmly established in previous studies, where cooperation can occur in a range of domains, from e.g. resource (time/frequency/power) control \cite{gesbert2007adaptation}, to e.g. beam alignment \cite{maschietti2017robust}.
	At the same time, the necessity of performing synchronous and mutually consistent decisions is challenging, especially in emerging heterogeneous, decentralized, network deployments where more and more decision need to be taken at the network's edge for the purpose of reducing latency \cite{mao2017survey}. What makes such scenarios difficult to deal with is the fact that channel feedback (based on which resource control decision are taken) cannot be centralized via the cloud and is often unreliable. For instance, devices will be endowed with different noisy estimates of the true channel state information (CSI).  In the past years a good amount of effort has been invested in deriving decentralized and noise-robust coordination policies while, only recently, the data-driven approach has been explored as a potential way to derive such coordination strategies in an automated way. In \cite{de2018team}, the problem of decentralized link scheduling with noisy CSI at the transmitter (CSIT) is formulated as a team decision problem and is solved with the so-called Team Deep Neural Networks (Team-DNNs). Team-DNNs are  jointly trained as “cooperative machines" maximizing the sum-rate of the system, yielding decentralized decision policies in the form of neural networks that prove to be robust to uncertainty in the local information, outperforming other conventional scheduling algorithms. In \cite{de2018robust}, Team-DNNs are considered to address the problem of decentralized precoding and, in spite of the challenge of outputting decision in a continuous space of precoding matrices, are shown to outperform state-of-the-art methods. In \cite{kim2018learning}, the Team-DNNs based transmitters are augmented with a supplementary neural network, designed to build succinct signaling information to be exchanged prior to taking actions in order to further increase coordination capabilities while maintaining moderated the size of messages exchanged.

	While Team-DNNs can be trained to reach robust cooperative decisions under arbitrary feedback noise, they are in principle trained for a fix uncertainty scenario (for instance fixed values of feedback noise levels). As a result, a limitation of this approach comes from the necessity of retraining the neural network ensemble whenever the feedback scenario evolves, i.e. whenever the feedback noise statistics change, in a wireless network. This drawback is exacerbated by the non-stationary nature of the wireless environment and the centralized nature of the training phase, rendering retraining not only potentially frequent but also burdensome. Instead, a machine learning solution able to yield robust cooperative decisions regardless of the time variations in the feedback quality would be highly desirable.

	In this work, we consider the data-driven policy design paradigm applied to the problem of distributed power control with noisy CSIT and we propose an approach to mitigate the retraining issue associated with Team-DNNs. Motivated by the fact that a single offline training phase is often preferable to multiple  run-time retraining phases impairing the performance, we introduce the Deep Mixture of Experts (DMoE) model in order to obtain a “universal" power control policy, able to adapt to a wide range of channel feedback quality scenarios.
	
	Deep Mixture of Experts model have been proposed in several prior works \cite{eigen2013learning,wang2018deep,shazeer2017outrageously,chazan2017speech} although, to the best of our knowledge, never in the context of wireless networking. Key background information on DMoE is given in Section \ref{sec:DMoE general}.  The key intuition behind DMoE models is akin to the issue of solving a complex problem combining experts specialized to deal with a particular subset of problem instances. The relevance of this approach to our machine learning problem is made explicit in Section \ref{sec:DMoE specific}.

	\section{Data-driven policy design for Team Decision problems}
	A team decision optimization problem arises whenever a group of decision makers seek to maximize a common utility by their actions, taken on the basis of local observations of the system state. It is formally defined as follows: 
	\begin{itemize}                                                                                                                                                                                                                                                                                                                                                                                                                                                                                  
		\item $K$: the number of decision makers (DMs). 
		\item $\pmb{s}\in \mathcal{S}$: the state of the world.
		\item $\hat{\pmb{s}}_{j}\in \hat{\mathcal{S}}_{j}$: the local information of user $j$.
		\item $\pi_j:\hat{\mathcal{S}}_j\rightarrow \mathcal{A}_j$: the decision policy of user $j$, mapping local observations into actions.
		\item $f:\mathcal{S}\times \prod_{j=1}^{K}\mathcal{A}_j\rightarrow \mathbb{R}$: a common utility, function of the state of the world and the actions of the DMs.
		\item $P_{s,\hat{s}_{1},\dots,\hat{s}_{K}}$: the joint distribution governing the relation between the state of the world and the local observations at the DMs.
	\end{itemize}
	A Team Decision (TD) solution is a set of decision policies that maximizes the expected utility and is the result of the following functional optimization problem 
	\begin{equation}
	(\pi_1^*,\dots,\pi_K^*)=\argmax_{\pi_1,\dots,\pi_K} \mathbb{E}\left[f(\pmb{s},\pi_1(\hat{\pmb{s}}_1),\dots,\pi_K(\hat{\pmb{s}}_K))\right]
	\label{TDsol}
	\end{equation}
	where the expectation is taken with respect to the random variables in bold.
	
	The distributed nature of the information is one of the distinctive traits in TD problems, in fact each decision maker $j$ is endowed with a local observation $\hat{\pmb{s}}_{j}$ that discloses only partial information about the world state $\pmb{s}$. For instance, in a wireless network design problem $\pmb{s}$ may represent the global channel state information matrix. Note that the way $\hat{\pmb{s}}_{j}$ is related to $\pmb{s}$ is very general and encompasses a range of practical situation, such as local feedback, noisy global feedback, hierarchical feedback, etc.\cite{gesbert2018team}. Also note that the optimization variables in (\ref{TDsol}) lie in the space of functions. Functional optimization problems are notoriously difficult to tackle and in order to circumvent them it is customary to represent each policy $\pi_i$ by a parametrized function $\pi_i^{\theta_i}$, recasting the original problem into the following 
	\begin{equation}
	(\theta_1^*,\dots,\theta_K^*)=\argmax_{\theta_1,\dots,\theta_K} \mathbb{E}\left[f(\pmb{s},\pi_1^{\theta_1}(\hat{\pmb{s}}_1),\dots,\pi_K^{\theta_K}(\hat{\pmb{s}}_K))\right]
	\label{TDsolParam}
	\end{equation}
	\subsection{Team-DNNs}
	A particular choice of parametrized policy  is that offered by the output of a DNN parametrized by $\theta_i$. In this case the policies are realized by Team-DNNs that work cooperatively  so as to solve the maximization problem in (\ref{TDsolParam}). This allows exploiting their approximation power and the efficient parameters optimization algorithms available (e.g. back-propagation), leading to a fully data-driven procedure to design decision policies  \cite{lee2019deep,  kim2018learning,de2018team,de2018robust}. Namely, given  a training set $\mathcal{D}=\{(s,\hat{s}_1,\dots,\hat{s}_K)_i\}^n_{i=1}\sim P_{s,\hat{s}_{1},\dots,\hat{s}_{K}}^{\otimes n}$, the neural networks can be trained using gradient ascent with the following objective function 
	\begin{equation}
	\mathcal{U}(\theta_1,\dots,\theta_K)=\smashoperator{\sum_{(s,\hat{s}_1,\dots,\hat{s}_K)\in\mathcal{D}}}\frac{f(s,\pi_1^{\theta_1}(\hat{s}_1),\dots,\pi_K^{\theta_K}(\hat{s}_K))}{|\mathcal{D}|}
	\end{equation}
	However, it is important to notice that the distributed information model abstains DMs from accessing the gradient information, which depends on the true state of the world and on the actions of the other DMs. As a result, the training phase has to be centralized with perfect information sharing, temporary violating the original decentralized information model and leading to the so called “centralized training/decentralized testing" paradigm.

	\subsection{Application to wireless networking problems}
	Team optimization problems emerge frequently in wireless networking. Groups of devices are usually required to coordinate based on noisy local observations of the system's state. An additional degree of difficulty associated to these scenarios comes from the fact that the levels of uncertainty (or noise) in the local estimates are linked to time-varying processes such as mobility, devices positions, etc. On the other hand it is realistic to assume that some statistical information about the feedback noise in the local observations can be given to the various DMs by means of regular probing. One issue is faced when the feedback noise statistics vary in time. The Team-DNNs optimization requires in fact burdensome centralized retraining whenever the uncertainty levels at the DMs changes. Therefore, it becomes desirable to devise a “universal" machine learning model, trained once on a variety of uncertainty scenarios, being able to adapt  to future noise configurations.
	
	Let us assume that the joint distribution governing the relation between the $\pmb{s}$ and $\{\pmb{s}_i\}_{i=1}^K$ at the DMs can itself be parametrized by $\sigma$, a set of parameters linked to the feedback noise at the various DMs, namely
	$P_{s,\hat{s}_{1},\dots,\hat{s}_{K}|\sigma}$
	for $\sigma\in \Sigma$.
	We further assume that the network can provide all DMs with $\hat{\sigma}$, an estimate of $\sigma$, enriching the available local information. For instance, $\sigma$ could be a vector containing the feedback noise level across all transmitters, which may or may not be identical from transmitter to transmitter. In general, the estimated $\hat{\sigma}$ can be inaccurate and therefore we model it as random variable distributed according to $P_{\hat{\sigma}|\sigma}$. Nonetheless, the final objective of the TD problem remains designing policies maximizing the expected utility as in (\ref{TDsol}), where now the local observation at user $i$ is  $(\hat{\pmb{s}}_{i},\hat{\pmb{\sigma}})$ and the average utility is taken w.r.t. the aggregated distribution $P_{s,\hat{s}_{1},\dots,\hat{s}_{K}|\sigma}P_{\hat{\sigma}|\sigma}P_{\sigma}$, where $P_{\sigma}$ represents the prior distribution of the information noise statistic and can be assumed distributed uniformly in $\Sigma$ when it is unknown.
	The core of the optimization problem is then unchanged and, theoretically, the same data-driven approach can be used to carry out a single centralized training phase using a training set $\mathcal{D'}=\{(\sigma,\hat{\sigma},\hat{s}_1,\dots,\hat{s}_K)_i\}^n_{i=1}\sim  P_{s,\hat{s}_{1},\dots,\hat{s}_{K}|\sigma}P_{\hat{\sigma}|\sigma}P_{\sigma}^{\otimes n}$. Ideally, the outcome of this process would be a set of DNNs, representing the policies at the various DMs, being able to well-approximate the optimal distributed strategies for various information noise configurations and capable to adapt their joint behaviour based on the network estimate $\hat{\sigma}$ during testing time. Practically speaking, even if the neural network capacity can be enlarged by increasing the model size, it becomes critical to train a unique multi-layer network to mimic different coordination strategies on different occasions. For this reason, we propose the use of the so-called Mixture of Experts model to represent the policies at the decision makers.
	
	\subsection{Deep Mixture of Experts}
	\label{sec:DMoE general}
	Dating back to 1991, the Mixture of Expert (MoE) model has been proposed by Jacobs et al. \cite{jacobs1991adaptive} as an ensemble method based on the “dividi et impera" principle. According to this paradigm, a hard problem is first decomposed into simpler ones until solutions can be obtained for any of the sub-problems; then, the yielded results are recombined together to obtain the original solution.
	
	The MoE structure reflects this principle incorporating a gating function $g$, learning a partition of the problem input space, and a set of simple, easily trainable models $\{f_i\}^{n_e}_{i=1}$, specializing over these subspaces. For each input instance, the gating network outputs a probability mass distribution over experts' indices, which is used to weight their outputs and to obtain the final prediction.

	The Deep MoE (DMoE) is the extension of the MoE with single layer neural network experts to the multi-layer model. First suggested in \cite{eigen2013learning}  by concatenating two single layer feed-forward MoE, it has been extended to convolutional layers in \cite{wang2018deep}  and neural networks with memory units in \cite{shazeer2017outrageously}. 
	In \cite{chazan2017speech} both experts and gating were directly replaced by multi-layer neural networks.
	
	The DMoE's ability to specialize DNNs for different data regimes prompts its application to the coordination problem in wireless settings, in which heterogeneous coordination strategies may arise for different uncertainty levels in the local observation.
	
	\section{Application to Decentralized Power Control}
	\subsection{Problem Formulation}
	The problem of decentralized power control in interference channels with noisy CSIT can be formulated as a TD problem in the sense (\ref{TDsolParam}), as seen below.\\ Consider a $K$-user interference channel with single-antenna transmitters (TXs) and receivers (RXs), in which TX $i$ serves RX $i$ with a maximum transmit power $P_{max}$. The decision makers are the $K$ TX and the channel gain matrix $\pmb{G}\in\mathbb{R}^{K\times K}\sim P_{G}$ is the system state. The local observation at TX $i$  is the pair $(\hat{\pmb{G}_i},\hat{\pmb{\sigma}})$., where $\hat{\pmb{G}_i}$ is a noisy estimate of channel state $\pmb{G}$ and $\hat{\pmb{\sigma}}$ is a network estimate of the statistic of the gain feedback noise at the various TXs.
	
	A typical utility function is the sum-rate of the system which, under the assumption of Gaussian distributed with zero mean and unit variance information symbols and noise, can be expressed as \cite{cover2012elements}
	\begin{equation*}
	R(G,P_1,\dots,P_K)=\sum_{i=1}^{K}\log_2\left(1+\frac{G_{i,i}P_i}{1+\sum_{j\neq i}G_{j,i}P_j}\right)
	\end{equation*}
	Therefore the team decision problem consists in devising a set of power control policies $\pi_1,\dots,\pi_K$ 
	\[
	\pi_i: \left(\hat{{G}_i},\hat{\sigma}\right) \rightarrow p_i\in [0,P_{max}]
	\]
	maximizing 
	\[
	\mathbb{E}\left[R\left(\pmb{G},\pi_1(\hat{\pmb{G}}_1,\hat{\pmb{\sigma}}),\dots,\pi_K(\hat{\pmb{G}}_K,\hat{\pmb{\sigma}})\right)\right]
	\]
	\subsection{Mixture of Experts policy design}
	\label{sec:DMoE specific}
	\begin{figure}
		\centering
		\begin{tikzpicture}[auto,node distance=3cm]
		\tikzset{
			block/.style= {draw, rectangle, minimum height=1.5em,minimum width=3em},
			blockB/.style= {draw, rectangle, minimum height=1cm,minimum width=4cm},
			sum/.style = {draw, circle, node distance=2cm},
			input/.style  = {coordinate},  
			output/.style = {coordinate}
		}
		\node () at (4, 2.5)(gat) {\includegraphics[width=1cm]{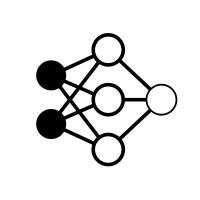}};
		\node () at (4,1.5) (envLab) {\scriptsize Experts};
		\node () at (4,3.25) (envLab) {\scriptsize Gating Net};
		\node () at (4, 0.75)(DM1) {\includegraphics[width=1cm]{DNN.png}};
		\node () at (4, 0) (n2) {$\vdots$};
		\node () at (4, -0.75)(DMK) {\includegraphics[width=1cm]{DNN.png}};
		\path [draw,-latex]	 (2,0.75) -- node [above] {$\hat{G}_{i},\hat{\sigma}$} (DM1.180);
		\path [draw,-latex] (2,-0.75) -- node [below] {$\hat{G}_{i},\hat{\sigma}$} (DMK.180);
		\node [sum,minimum height=2.4em] at (7, 0)(ut) {$\sum$};
		\node [sum, inner sep=0pt] at (5.5, 0.75)(m1) {$\times$};
		\node [sum, inner sep=0pt] at (6, - 0.75)(m2) {$\times$};
		\path [draw,-latex]	 (DM1.0) -- node [above] {$p_{j,1}$} (m1.180);
		\path [draw,-latex]	 (m1.0) --  (ut.160);
		\path [draw,-latex]	 (DMK.0) -- node [above] {$p_{j,n_e}$} (m2.180);
		\path [draw,-latex]	 (m2.0) --  (ut.200);
		\path [draw,-latex] (3,2.5) -- node [above] {$\hat{\sigma}$} (gat);
		\path [draw,-latex] (gat)-- (5.5, 2.5) --  (m1.90);
		\path [draw,-latex] (5.5, 2.5) -- (6, 2.5) -- (m2.90);
		\path [draw,-latex] (ut)-- node [above] {$p_j$} (8, 0);
		\end{tikzpicture}
		\caption{The structure of the Mixture of Expert used to represent policies at the DMs}
		\vspace*{-1em}
		\label{DMOE}
	\end{figure}
	In this distributed power control setting with  fixed CSI quality at the TXs, it has already been shown that the data-driven policy design with multi-layer neural networks can give coordination strategies that are robust to uncertainty in the local observations and that are outperforming model-based algorithms such as WMMSE \cite{kim2018learning}. Moreover, it has been noticed that for different uncertainty level at the TXs, the yielded policies exhibits distinct behaviours.
	We aim at capturing this heterogeneous spectrum of power control algorithms by means of an adaptable model exploiting the above-introduced data-driven approach and DMoEs.
	
	Specifically, the power control algorithm at each DM $j$ is obtained learning a set of experts functions $\{f^{\theta_{k,j}}\}_{k=1}^{n_e}$ along with a gating function $g^{\theta_{j}}$.
	Each expert function at DM $j$ is of the form
	\[
	f^{\theta_{k,j}}:\left(\hat{{G}_i},\hat{\sigma}\right)\rightarrow p_{j,k}\in  [0,P_{max}], \quad  \text{for }k\in {1,\dots,n_e}
	\]
	and is realized by a neural network of parameters $\theta_{k,j}$.
	Similarly, the gating function is implemented by a neural network of parameters $\theta_{j}$ and is of the form 
	\[
	g^{\theta_{j}}:\left(\hat{\sigma}\right)\rightarrow  w\in \mathbb{R}^{n_e}
	\]
	where $w$ is a vector in the $n_e-1$ dimensional simplex.\\
	The final power policy at DM $j$ is a combination of these DNNs outputs and is given by
	\[
	\pi_j(\hat{{G}_i},\hat{\sigma})=\sum_{k=1}^{n_e}f^{\theta_{k,j}}(\hat{{G}_i},\hat{\sigma})g_k^{\theta_{j}}(\hat{\sigma})
	\]
	where the $g_k^{\theta_{j}}(\hat{\sigma})$ is the $k$-th component of vector $g^{\theta_{j}}(\hat{\sigma})$.

	\section{Experiments}
	We consider a two-user interference channel for the ease of displaying results, these can be extended to more transmitters resizing the Team-DMoE model in order to cope with the increased dimension of the input variables.

	In our simulation we consider a Rayleigh fading channel such that the entries of the channel gain matrix $\pmb{G}\in\mathbb{R}^{2\times 2}$ are independent chi-squared random variables. We assume that the CSI available at user $i$ is distributed according to
	\[
	\hat{\pmb{G}_i}=\Gamma_{i}\odot \pmb{G} +\sqrt{1-\Gamma^2_{i}}\odot\pmb{\Delta}^{(i)}
	\]
	where $\odot$ denotes the element-wise product, $\pmb{\Delta}^{(i)}\in\mathbb{R}^{2\times 2}$ has i.i.d. chi-squared entries and $\Gamma_{i}\in[0,1]$ represents the degree of uncertainty at TX $i$.
	The estimate of the feedback noise parameters $\pmb{\sigma}=[\Gamma_{1},\Gamma_{2}]$ is given by
	\[
	\hat{\pmb{\sigma}}=\pmb{\sigma}+\pmb{\Delta}=[\Gamma_{1},\Gamma_{2}]+\pmb{\Delta}
	\]
	where $\pmb{\Delta}\in \mathbb{R}^{2}$ is a random vector with i.i.d. Gaussian entries with zero mean and variance $\sigma_n$.

	We use TensorFlow to implement and train the Team-DMoE model. Both DMoEs comprise two identical experts realized by a three-layer DNNs, each layer has 10 neurons  with ReLU activation functions while at the output layer we use Sigmoid activations multiplied by $P_{max}$ in order to satisfy the power constraint. The gating network is a two-hidden layer DNN with 10 neurons and ReLu activation functions, we use Softmax at the output layer to select the best expert for each input.
	We use a training set of 100000 samples and we perform 8000 gradient updates using batches of 1000 samples alternating between expert and gating network optimization. During the former phase, we directly optimize the sum-rate utility feeding experts with training samples according to the gating output assignment. During the latter, we train the gating net to assign each training sample to the best performing expert.
	As noted in \cite{de2018team}, a proper initialization is necessary in order to have convergence to good local minima, therefore we  initialize a pair of experts for the low uncertainty regime ($[\Gamma_{1},\Gamma_{2}]\approx [0,0]$) and the other for the high uncertainty regime ($[\Gamma_{1},\Gamma_{2}]\approx [1,1]$).
	
\begin{figure}
		\centering
		\includegraphics[width=0.9\columnwidth]{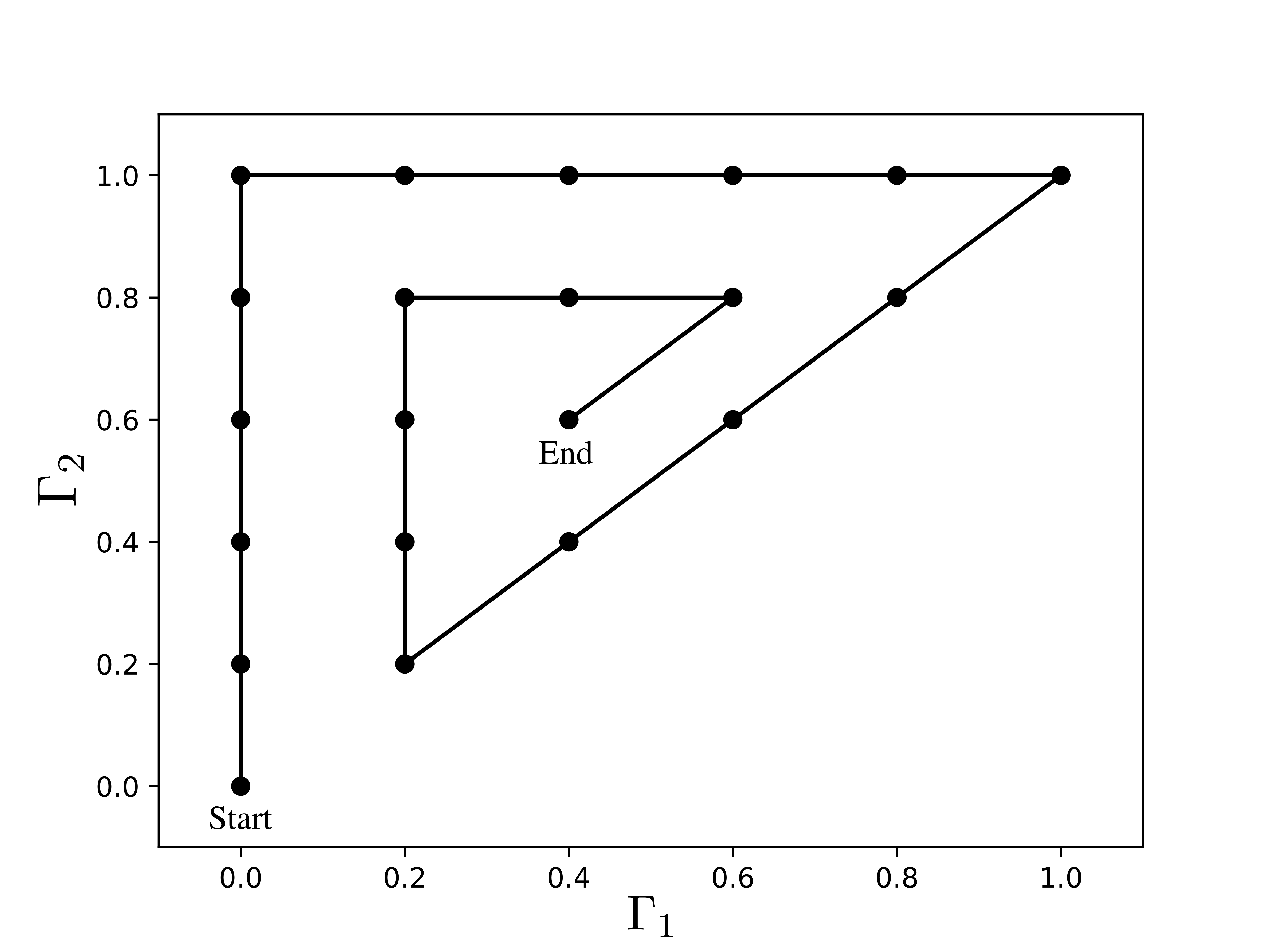}
		\caption{Trajectory of the feedback noise parameters $(\Gamma_1,\Gamma_2)$.}
		\label{fig:Uncertainty}
		\vspace*{-1em}
	\end{figure}

	As benchmark scenario, we consider a time-slotted communication system with the channel noise statistic changing every 10 time slots. The uncertainty levels $[\Gamma_{1},\Gamma_{2}]$ follow the trajectory depicted in Fig \ref{fig:Uncertainty}, starting from the perfect CSI setting. The trajectory is chosen to span a variety of the noise scenarios allowing to assess the performance of Team-DMoEs in various uncertainty regimes. For instance, at the start of the experiment, both TXs start with perfect channel gain information. By the 6th time interval, TX 2 has lost all its channel gain feedback. By the 11th time interval, both TXs have lost all channel gain information, before gradually recovering it, etc.
	
	\begin{figure*}
		\centering
		\begin{subfigure}[b]{0.9\columnwidth}
			\centering
			\includegraphics[width=\columnwidth]{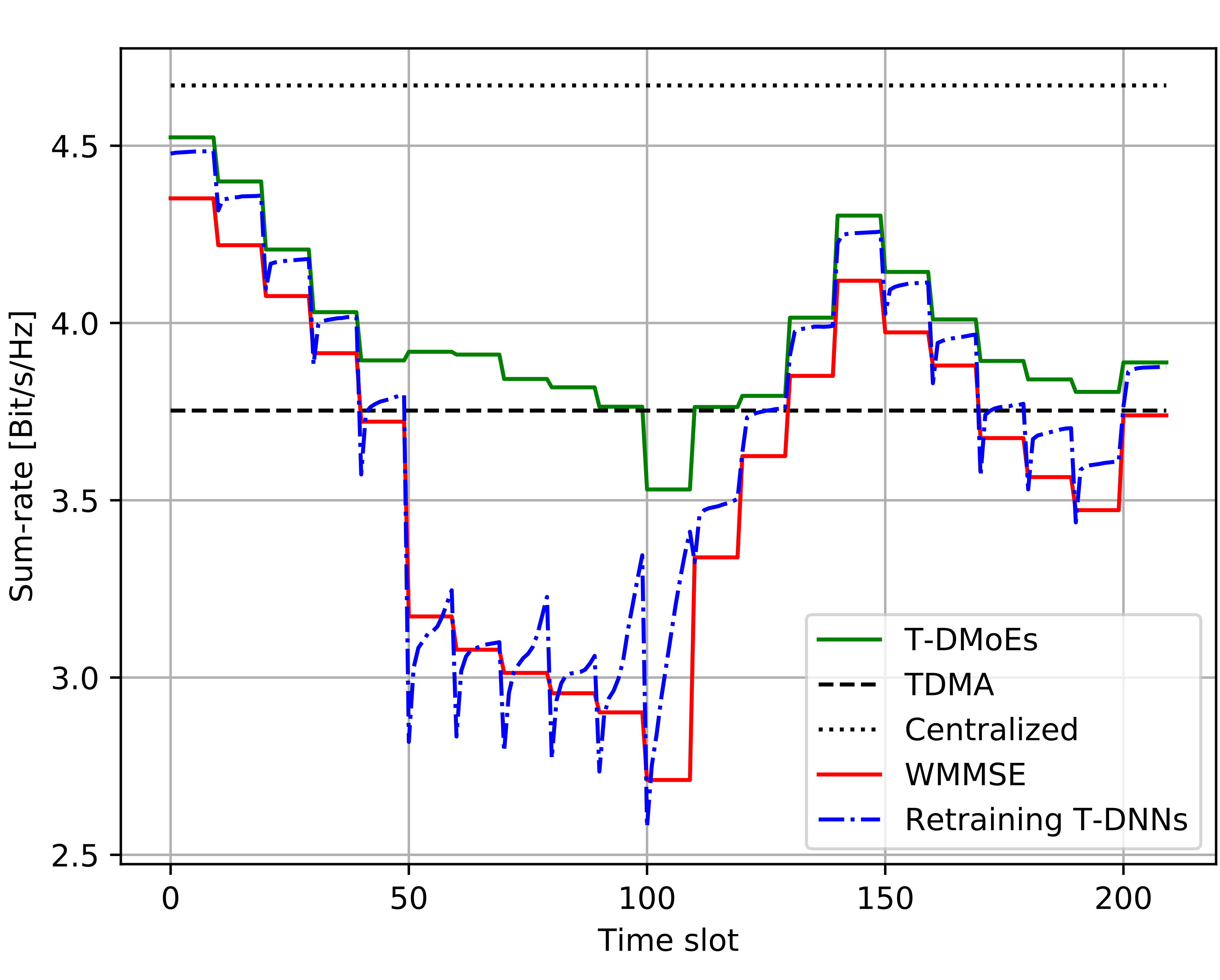}
			\caption{$R_{up}=10$}
			\label{fig:Ru10}
		\end{subfigure}
		\hspace{2.9em}
		\begin{subfigure}[b]{0.9\columnwidth}
			\centering
			\includegraphics[width=\columnwidth]{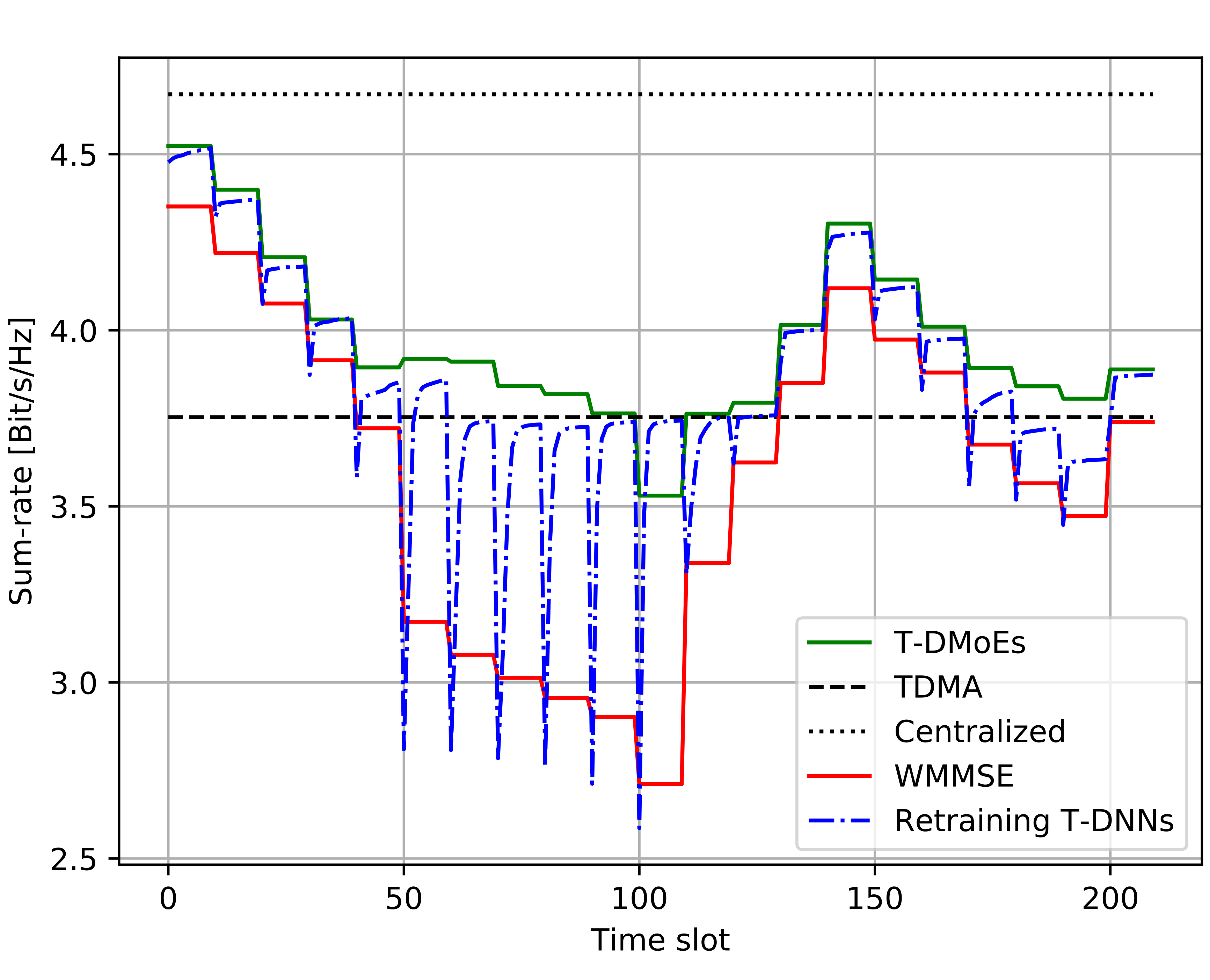}
			\caption{$R_{up}=100$}
			\label{fig:Ru100}
		\end{subfigure}
		\caption{Average sum-rate of the various power control algorithms}
		\label{fig:Adaptability}
		\vspace*{-1em}
	\end{figure*}
	\begin{figure}
		\vspace*{-1.4em}
		\centering
		\includegraphics[width=0.9\columnwidth]{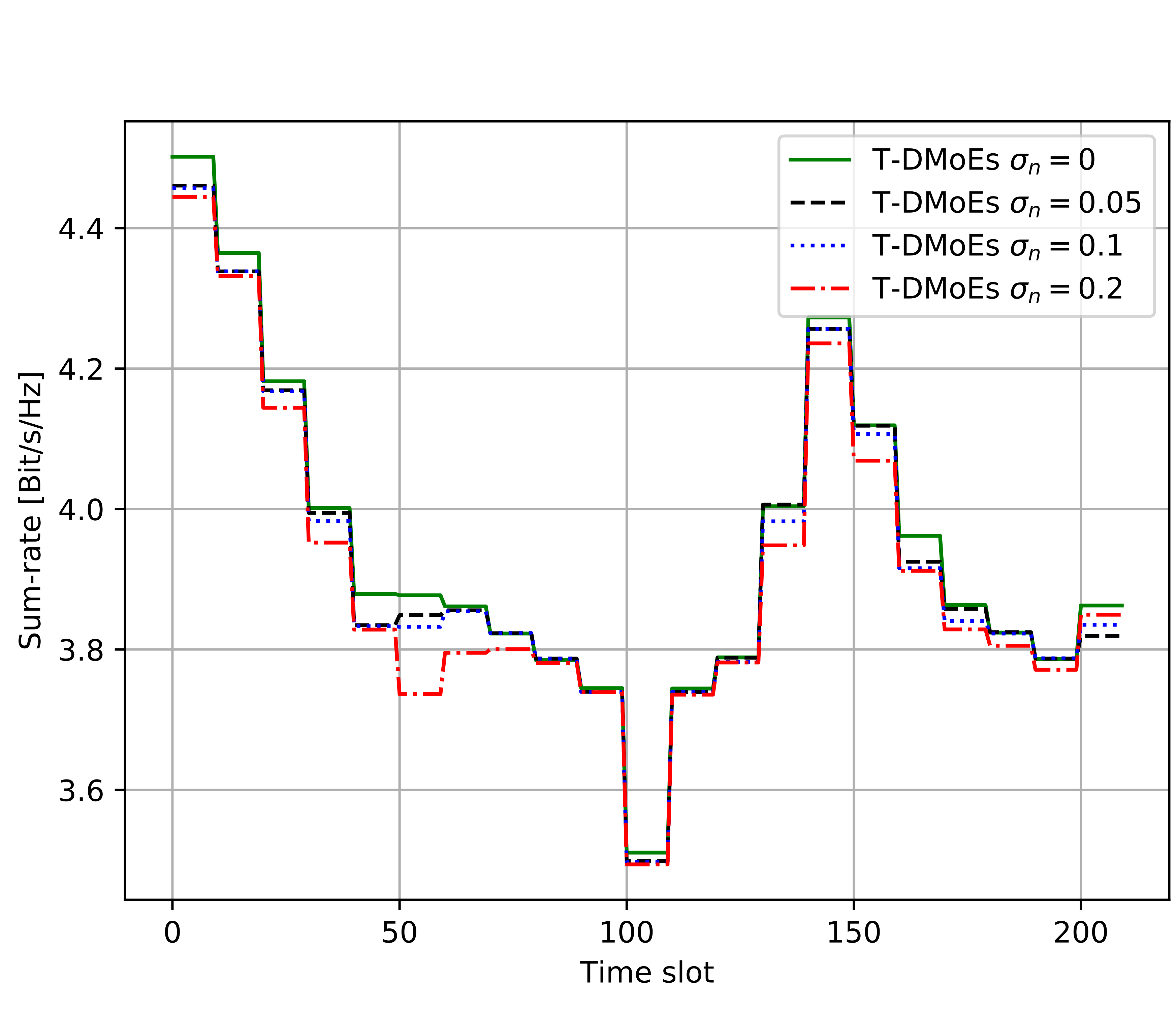}
		\caption{T-DMoEs sum-rate for different estimation noise.}
		\label{fig:NoisyEstimates}
		\vspace*{-1em}
	\end{figure} 
	We compare the performance of the Team-DMoEs model with the following power control schemes:
	\begin{itemize}
		\item \emph{Perfect-CSI:} both TX have perfect CSI knowledge and use the optimal power control strategy.
		\item \emph{Informationally Naive WMMSE:} both TX use the WMMSE power algorithm of \cite{shi2011iteratively} assuming their local info being perfect.
		\item \emph{TDMA:} Only one of the transmitters is allowed to transmit with power $P_{max}$.
		\item \emph{Team-DNNs:} A pair of DNNs that are constantly retraining in order to derive the coordination policy for the current uncertainty levels. For every noise configuration, the training starts from the WMMSE solution and is carried out on a training set of 30000 samples coming from the current joint distribution and performing $R_{up}$ gradient updates in every time slot with a batch of size 1000. 
	\end{itemize}
	
	For every transmission slot, we compute the average sum-rate over 10000 channel realization. In Fig. \ref{fig:Ru10}, the Team-DMoE with perfect estimation $(\sigma_n=0)$ delivers the highest sum-rate in almost all uncertainty regimes, the only exception is in the case where both TXs have poor CSI knowledge in which case TDMA performs better being the optimal power control scheme for that scenario. On the other hand, the constantly trained T-DNN, being initialized to the WMMSE, delivers good performance only when the information at DM is not too degraded, in all other cases the training process impairs the average sum-rate. Retraining T-DNNs perform the worst in correspondence of the changes in the noise levels that prompt the retraining process; moreover, it can be noticed that given a computational power that allows 10 gradient updates on each time slot ($R_{up}=10$) the Team-DNNs are not able to converge to a good power control policy before channel switches to the next noise configuration. For this reason in Fig.\ref{fig:Ru100} we decide to provide the Team-DNNs with 10 time the computational power, letting them to perform 100 updates every time slots. In this case the overall number of gradient updates is just sufficient to converge to a good local model in most of the cases.
	
	In Fig. \ref{fig:NoisyEstimates} we evaluate the performance of the Team-DMoEs for different values of $\sigma_n$. The performance reduction due to imperfect estimates degrades gracefully as the value of $\sigma_n$ increases, indicating that the derived policies are robust to such impairment.

	\section{Conclusion}
	We focused on how retraining overhead can be mitigated in the context of coordination problems with variable uncertainty levels at the DMs. Specifically, we proposed a network architecture inspired by the DMoE model that exploits the estimate of feedback noise statistics to adapt its behavior and to bypass retraining phases. We then evaluated its performance in the problem of distributed power control, showing that it outperforms other power control algorithms across most of the noise level configurations. Because the problem  of training overhead extends to additional ML applications in wireless communications \cite{simeone2020learning,park2019meta}, it is of primary importance to further investigate what are the best network architectures and training algorithms to reduce it.
	\section*{Acknowledgement} 
The work of M.Zecchin is founded by Marie Skłodowska-Curie actions
	(MSCA-ITN-ETN 813999 WINDMILL).
	\bibliography{SPAWCMoE} 

\begin{thebibliography}{10}

\bibitem{gesbert2007adaptation}
D.~Gesbert, S.~G. Kiani, A.~Gjendemsjo, and G.~E. Oien, ``Adaptation,
  coordination, and distributed resource allocation in interference-limited
  wireless networks,'' {\em Proceedings of the IEEE}, vol.~95, no.~12,
  pp.~2393--2409, 2007.

\bibitem{maschietti2017robust}
F.~Maschietti, D.~Gesbert, P.~de~Kerret, and H.~Wymeersch, ``Robust
  location-aided beam alignment in millimeter wave massive mimo,'' in {\em
  GLOBECOM 2017-2017 IEEE Global Communications Conference}, pp.~1--6, IEEE,
  2017.

\bibitem{mao2017survey}
Y.~Mao, C.~You, J.~Zhang, K.~Huang, and K.~B. Letaief, ``A survey on mobile
  edge computing: The communication perspective,'' {\em IEEE Communications
  Surveys \& Tutorials}, vol.~19, no.~4, pp.~2322--2358, 2017.

\bibitem{de2018team}
P.~de~Kerret, D.~Gesbert, and M.~Filippone, ``Team deep neural networks for
  interference channels,'' in {\em 2018 IEEE International Conference on
  Communications Workshops (ICC Workshops)}, pp.~1--6, IEEE, 2018.

\bibitem{de2018robust}
P.~de~Kerret and D.~Gesbert, ``Robust decentralized joint precoding using team
  deep neural network,'' in {\em 2018 15th International Symposium on Wireless
  Communication Systems (ISWCS)}, pp.~1--5, IEEE, 2018.

\bibitem{kim2018learning}
M.~Kim, P.~de~Kerret, and D.~Gesbert, ``Learning to cooperate in decentralized
  wireless networks,'' in {\em 2018 52nd Asilomar Conference on Signals,
  Systems, and Computers}, pp.~281--285, IEEE, 2018.

\bibitem{eigen2013learning}
D.~Eigen, M.~Ranzato, and I.~Sutskever, ``Learning factored representations in
  a deep mixture of experts,'' {\em arXiv preprint arXiv:1312.4314}, 2013.

\bibitem{wang2018deep}
X.~Wang, F.~Yu, L.~Dunlap, Y.-A. Ma, R.~Wang, A.~Mirhoseini, T.~Darrell, and
  J.~E. Gonzalez, ``Deep mixture of experts via shallow embedding,'' {\em arXiv
  preprint arXiv:1806.01531}, 2018.

\bibitem{shazeer2017outrageously}
N.~Shazeer, A.~Mirhoseini, K.~Maziarz, A.~Davis, Q.~Le, G.~Hinton, and J.~Dean,
  ``Outrageously large neural networks: The sparsely-gated mixture-of-experts
  layer,'' {\em arXiv preprint arXiv:1701.06538}, 2017.

\bibitem{chazan2017speech}
S.~E. Chazan, J.~Goldberger, and S.~Gannot, ``Speech enhancement using a deep
  mixture of experts,'' {\em arXiv preprint arXiv:1703.09302}, 2017.

\bibitem{gesbert2018team}
D.~Gesbert and P.~de~Kerret, ``Team methods for device cooperation in wireless
  networks,'' in {\em Cooperative and Graph Signal Processing}, pp.~469--487,
  Elsevier, 2018.

\bibitem{lee2019deep}
H.~Lee, S.~H. Lee, and T.~Q. Quek, ``Deep learning for distributed
  optimization: Applications to wireless resource management,'' {\em arXiv
  preprint arXiv:1905.13378}, 2019.

\bibitem{jacobs1991adaptive}
R.~A. Jacobs, M.~I. Jordan, S.~J. Nowlan, G.~E. Hinton, {\em et~al.},
  ``Adaptive mixtures of local experts.,'' {\em Neural computation}, vol.~3,
  no.~1, pp.~79--87, 1991.

\bibitem{cover2012elements}
T.~M. Cover and J.~A. Thomas, {\em Elements of information theory}.
\newblock John Wiley \& Sons, 2012.

\bibitem{shi2011iteratively}
Q.~Shi, M.~Razaviyayn, Z.-Q. Luo, and C.~He, ``An iteratively weighted mmse
  approach to distributed sum-utility maximization for a mimo interfering
  broadcast channel,'' {\em IEEE Transactions on Signal Processing}, vol.~59,
  no.~9, pp.~4331--4340, 2011.

\bibitem{simeone2020learning}
O.~Simeone, S.~Park, and J.~Kang, ``From learning to meta-learning: Reduced
  training overhead and complexity for communication systems,'' 2020.

\bibitem{park2019meta}
S.~Park, O.~Simeone, and J.~Kang, ``Meta-learning to communicate: Fast
  end-to-end training for fading channels,'' {\em arXiv preprint
  arXiv:1910.09945}, 2019.

\end{thebibliography}
	\bibliographystyle{ieeetr}
	
\end{document}